\documentclass[12pt]{iopart}
\begin{document}

\title[Lorentz Invariance and the semiclassical approximation of loop quantum gravity]{Lorentz Invariance and
the semiclassical approximation of loop quantum gravity.}

\author{Carlos N Kozameh and Florencia Parisi}

\address{Facultad de Matem\'atica, Astronom\'\i a y F\'\i sica, \\ Universidad
Nacional de C\'ordoba, \\ Ciudad Universitaria, (5000) C\'ordoba, Argentina}

\begin{abstract}

It is shown that the field equations derived from an effective interaction hamiltonian for Maxwell and
gravitational fields in the semiclassical approximation of loop quantum gravity using rotational invariant
states (such as weave states) are Lorentz invariant. To derive this result, which is in agreement with the
observational evidence, we use the geometrical properties of the electromagnetic field.

\end{abstract}

 \ead{kozameh@fis.uncor.edu} \maketitle

\section{Introduction}

In recent years there has been hope of observing quantum gravity effects via the propagation of light through
cosmological distances. This hope is based in some models describing the interaction of quantum Maxwell and
gravitational fields that predict a breakdown of Lorentz invariance at a linearized level in the semiclassical
approximation. The common feature in these models is a non standard dispersion relation which shows that
spacetime behaves as a medium with a frequency dependent index of refraction \cite{A-Camelia, GaPu, Alfaro,
Thiemann}.

There are two predicted effects that in principle can be measured with the present technology. One is the
dependence of the speed of photons on their energy. More energetic photons would travel slower than less
energetic ones and thus, there would be a time delay between them as they travel through a cosmological
distance \cite{A-Camelia, GaPu}. The second effect is an additional dependence of the speed of photons on their
helicity \cite{GaPu} (violating parity and Lorentz symmetry). As a consequence, the polarization direction can
rotate around $2\pi$ over an energy dependent distance (short compared to a typical cosmological distance). The
net effect on light coming from a cosmological source that emits synchrotron radiation would be a loss of
polarization by the time it reaches the observer \cite{GleiKo}.

However, recent observations of light emitted from synchrotron sources put severe limits on these models and
raise doubts on the validity of the linear approximations since the observations are consistent with Lorentz
invariance \cite{Jacobson1}. Moreover, applying the results of Gleiser- Kozameh \cite{GleiKo} to polarized
radiation from very energetic photons coming from gamma ray bursts \cite{GRB}, it is possible to place an
extremely small bound on the phenomenological parameter that corresponds to parity violation. The new bound
corresponding to $2$ MeV photons is given by $\chi< 10^{-15}$, the lowest obtained so far\cite{Jacobson2}. The
lack of evidence supporting Lorentz violation motivated a revision of the effective Hamiltonian describing the
interaction between the Maxwell and gravitational fields in the semiclassical approximation. Very recently it
was shown that, at least in the canonical framework, Lorentz invariance is preserved at a linearized level when
the geometrical meaning of the Maxwell field is taken into account \cite{GleiKoPa}.

It is worth pointing out that if Lorentz violation occurs as a second order effect, it would be impossible to be
measured with a time delay observation. The predicted order of magnitude for this measurement would change from
$10^{-3}$sec for a violation at a linear level in $E/E_{Planck}$ (for an $E=1 GeV$ photon) to $10^{-22}$sec for
a quadratic correction. However, it could in principle be detected by observing the rotation of the polarization
direction. Since this second prediction is a consequence of parity violation in the canonical approach to
quantum gravity, in this work we reexamine the effective Hamiltonian of the Gambini-Pullin model \cite{GaPu}
keeping all the orders in the Taylor expansion that naturally arises when the rotational invariant properties of
the weave states are used to compute expectation values. We show that the resulting equations preserve Lorentz
invariance to all orders if we impose the geometrical condition that the electromagnetic field operator is the
curvature of the quantum Maxwell connection. Although the calculations in this work neglect back reaction terms
in the metric, we argue in the concluding remarks that the full set of equations would preserve Lorentz
invariance at a local level.

\section{The Interaction Hamiltonian and the field equations}
In this section we first compute the effective Hamiltonian density for the interaction between the Maxwell and
gravitational field in loop quantum gravity. The calculation is done by taking expectation values of the energy
density operator using weave states for the gravitational sector, and coherent states for the Maxwell sector
\cite{GaPu}. We then derive the corresponding field equations and show that they can be written as the standard
Maxwell equations.

The Maxwell field on a curved spacetime is given by an exact 2-form
$$F= dA,$$
 with $A$ the Maxwell connection 1-form. Given a local coordinate system with
 timelike and spacelike coordinate vectors $e_o^{\alpha}$ and $e_a^{\alpha}$ respectively,
 we define the electric and magnetic fields as $E_a = F_{ao}$ and
 $B_{a} =\frac{1}{2}\epsilon _{abc}F_{bc}$, where $\epsilon_{abc}$ is the Levi-Civita symbol.

The energy density is:

\begin{equation}
\mathcal{H}=\frac{1}{2}g^{ab}\sqrt{-g}\left(E_aE_b+B_aB_b\right)
\end{equation}
where $g^{ab}$ is the 3-dimensional metric and $g$ is the determinant of the 4-dimensional metric.
In order to obtain the corresponding Hamiltonian density we must define the phase space.
We will take the pairs $([A_a],\pi^b)$ as members of our phase space,
where $[A_a]$ represents a class of equivalence
(with $\tilde{A_a}$ and $A_a$ being
equivalent if $\tilde{A_a}=A_a+\partial_a\phi$) and $\pi^b$ is such that $\partial_b\pi^b=0$ \cite{Wald}.
With the above definitions, $E_a$ and $B_a$ are given by
$E_a=-\partial_t[A_a]$ and $B_a=\epsilon_{abc}\partial_bA_c$.

Promoting the metric to a quantum operator as proposed by Thiemann \cite{Thiemann2}, the electric part of the
energy density operator is given by

\begin{equation}
\mathcal{\hat{H}}_E=\frac{1}{2}\int d^3x \int d^3y\hat w^a_k(x)\hat w^b_k(y)E_a(x)E_b(y)f_{\epsilon}(x-y),
\end{equation}
where $\hat w^a_k$ are the quantum operators corresponding to the vectors of a triad and
$f_{\epsilon}(x-y)$ is a function introduced for regularization purposes that goes to a delta function as
$\epsilon\rightarrow0$. Taking the expectation value in a weave state $|\Delta\rangle$, one gets

\begin{equation}
\label{expvalue}\langle\Delta|\mathcal{\hat{H}}_E|\Delta\rangle=\frac{1}{2}\sum_{v_i,v_j} \langle\Delta|\hat
w^a_k(v_i)\hat w^b_k(v_j)|\Delta\rangle E_a(v_i)E_b(v_j),
\end{equation}
since the operators $\hat w^a_k$ only act at the vertices $v_i$ of the state. The above summation must be
performed on vertices within a region of dimension $\Delta$ such that the energy density becomes a scalar under
spatial rotations. Assuming that the electric field is slowly varying in a $\Delta$ scale, we can expand it
around the central point $P$ as

\begin{equation}
E_a(v_i)=E_a(P)+\sum_k \frac{1}{k!}(v_i-P)^{c_1}...(v_i-P)^{c_k}\partial_{c_1}...\partial_{c_k}E_a(P),
\end{equation}
where the distance $(v_i-P)^{c_1}$ is assumed to be at least of order of the Planck length $\ell_P$.

Inserting the above expression in (\ref{expvalue}), the electric part of the effective energy density can be
written as

\begin{equation}
\label{H_E} \mathcal{H}_{E}=\frac{1}{2}\delta^{ab}E_aE_b + E_a \sum_{k=1}^\infty \ell_P^k
\theta_k^{abc_1...c_k}\partial_{c_1}...\partial_{c_k} E_b \equiv \frac{1}{2}E_aH^{ab}E_b,
\end{equation}
with $\theta_k^{abc_1...c_k}$ a field-independent tensor .

Since $\mathcal{H}_{E}$ is a scalar, the quantity $\theta_k^{abc_1...c_k}\partial_{c_1}...\partial_{c_k} E_b$
must transform as a vector under spatial rotations. Thus, it can be written as the curl of a vector field plus
the gradient of a scalar function:

\begin{equation}
\theta_k^{abc_1...c_k}\partial_{c_1}...\partial_{c_k} E_b=\left(\nabla\times\vec{C}_{1k}\right)^a +
\nabla^a\phi_{1k},
\end{equation}
but, again $\vec{C}_{1k}=\left(\nabla\times\vec{C}_{2k}\right) + \nabla \phi_{2k}$ and so on
(when needed, we will switch for simplicity between index and vectorial notation).  After $k$ steps
we have
\begin{equation}
\theta_k^{abc_1...c_k}\partial_{c_1}...\partial_{c_k} E_b=\alpha_k \left[\left(\nabla\times\right)^k
\vec{E} + \nabla \phi_k\right]^a,
\end{equation}
where the expression $\left(\nabla\times\right)^k$ must be interpreted as $k$ curls taken consecutively,
$\alpha_k$ are constant coefficients  and $\phi_k$ is a linear functional of $E_a$.

Introducing this into (\ref{H_E}) we obtain

\begin{equation}
\mathcal{H}_{E}=\frac{1}{2}\delta^{ab}E_aE_b + E_a \sum_{k=1}^\infty \ell_P^k \alpha_k \left[\left(\nabla\times\right)^k
\vec{E} + \nabla \phi_k\right]^a. \label{hamiltonian}
\end{equation}

From the above expression one writes the explicit action of $H^{ab}$ on $E_b$ as

\begin{equation}
H^{ab}E_b = E^a
 + 2 \sum_{k=1}^\infty \ell_P^k \alpha_k[(\nabla\times)^k\vec{E}]^a + \nabla^a \phi(\vec{E}),
\end{equation}
where $\phi(\vec{E})=\sum_{k=1}^\infty \ell_P^k \alpha_k \phi_k(\vec{E})$ and it is linear in the electric field.

In order to obtain the Hamiltonian density we must first determine the relationship between the electric field
$E_a$ and the canonical momentum $\pi^b$. We know that, by definition, $E_a=-\partial_t[A_a]$ and, from
Hamilton's equations, $\partial_t[A_a]= \frac{\partial\mathcal{H}}{\partial\pi^a}$, hence

\begin{equation}
\label{E_pi}E_a=-\frac{\partial\mathcal{H}}{\partial\pi^a}= -\frac{\partial\mathcal{H}}{\partial E_b}
\frac{\partial E_b}{\partial\pi^a}.
\end{equation}

Note that the above is a differential equation for $E_a(\pi^b)$. Inserting (\ref{hamiltonian}) into (\ref{E_pi}) gives
\begin{equation}
\left(\frac{\partial E_b}{\partial\pi^a} H^{bc}+\delta_a^c\right)E_c=0.
\end{equation}

 Solving this equation for $E_a$ yields

\begin{equation}
\label{E-pi}E_a=-(H^{-1})_{ab}\pi^b.
\end{equation}

This is the desired relationship between $E_a$ and $\pi^b$. Inserting Eq. (\ref{E-pi}) in the energy density
yields the hamiltonian density.

\begin{equation}
\mathcal{H}=\frac{1}{2}\left(\pi^a H^{-1}{}_{ab}\pi^b + B_a H^{ab}B_b\right)
\end{equation}

Since both $\pi^a$ and $B_a$ are divergence free fields, integrating by parts and dropping surface terms, it is
possible to redefine $H^{ab}$ to eliminate the gradient part of the operator. We thus write

\begin{equation}
\mathcal{H}=\frac{1}{2}\left(\vec{\pi}\textbf{\textsc{H}}^{-1}\vec{\pi} +
\vec{B}\textbf{\textsc{H}}\vec{B}\right)
\end{equation}

with
\begin{equation}
\textbf{\textsc{H}}=Id
+ 2 \sum_{k=1}^\infty \ell_P^k \alpha_k(\nabla\times)^k.
\end{equation}

We now derive the corresponding field equations for $\vec{E}$ and $\vec{B}$. The homogeneous Maxwell equations
\begin{eqnarray}
\nabla\cdot\vec{B}= 0 \\
\partial_t\vec{B} + \nabla\times\vec{E}= 0,
\end{eqnarray}
are automatically satisfied since $\vec{E}=-\partial_t\vec{A}$ and $\vec{B}=\nabla\times\vec{A}$.

Moreover, using $\vec{\pi}=-\textbf{\textsc{H}}\vec{E}$ we get

\begin{equation}
0 =\nabla\cdot\vec{\pi}= \nabla\cdot\vec{E},
\end{equation}
where the explicit form of $\textbf{\textsc{H}}$ has been used.

The remaining Hamilton equation $\partial_t\pi^a=-\frac{\partial\mathcal{H}}{\partial A_a}$ yields (using again
$\vec{\pi}=-\textbf{\textsc{H}}\vec{E}$),

\begin{equation}
\textbf{\textsc{H}}(\partial_t\vec{E})=\nabla\times(\textbf{\textsc{H}}\vec{B}).
\end{equation}

Since the curl operator commutes with $\textbf{\textsc{H}}$, the above equation reduces to

\begin{equation}
\partial_t\vec{E}=\nabla\times\vec{B},
\end{equation}
which proves that, within the semiclassical approximation the electromagnetic field satisfies the usual Maxwell
equations and it is therefore Lorentz invariant.

\section{Final comments and conclusions}
In this work we have shown that when we take into account the typical volume of a weave state to construct a
rotational invariant expectation value and assume that the Maxwell field admits a Taylor expansion (taking  the
Planck length as an expansion parameter), the resulting field equations are Lorentz invariant. An equivalent
statement would be that the discreteness introduced by the weaves do not modify the invariant properties of the
classical field.

Since only invariant properties of the energy density under spatial rotations have been used to derive these
results, Lorentz invariance should also remain valid for other rotationally invariant states as long as the same
regularization procedure is used to construct the energy density operator.

A second observation to be made is that $-\vec{E}$ is not the conjugate momentum to $\vec{A}$. Note also that
Eq.(\ref{E-pi}) is the quantum equivalent of $E_a = -g_{ab}\pi^b$ with the classical $g_{ab}$ replaced by
$H^{-1}{}_{ab}$. Therefore one should not extrapolate from the classical result and make the (wrong) assumption
that $-\vec{E}$ and $\vec{A}$ are conjugate pairs when quantum gravity effects are taken into account.

 One could argue that these states do not constitute an appropriate basis for computing expectation values
of second or higher order corrections to the semiclassical approximation and we do not know if indeed this is
the case. However, it seems plausible that one can always find a new basis such that, at each order in a
perturbed solution of the coupled field equations, the expectation value of the metric operator yields the
classical value of the metric field in the Einstein-Maxwell equations of motion resulting from a classical
hamiltonian. If there exists a one-parameter family of basis connecting the weave states with the new basis,
then it is straightforward to generalize the above results and prove Lorentz invariance for the full theory.

\ack

We thank Ted Jacobson for valuable help. This work was supported in part by grants of the National University of C\'ordoba and Agencia C\'ordoba
Ciencias. (Argentina). C.N.K. is a member of CONICET. F.P. is supported by SECYT - UNC.


\end{document}